\documentclass[10pt]{iopart}
\usepackage{graphicx}
\usepackage{epstopdf}
\usepackage{ulem} 
\begin{document}

\title{Comparative study of the two-phonon Raman bands of silicene and graphene}

\author{Valentin N. Popov$^1$ and Philippe Lambin$^2$}

\address{$^1$Faculty of Physics, University of Sofia, BG-1164 Sofia, Bulgaria}

\address{$^2$Research Center in Physics of Matter and Radiation, University of Namur (FUNDP), B-5000 Namur, Belgium}

\ead{vpopov@phys.uni-sofia.bg}

\begin{abstract}
We present a computational study of the two-phonon Raman spectra of silicene and graphene within a density-functional non-orthogonal tight-binding model. Due to the presence of linear bands close to the Fermi energy in the electronic structure of both structures, the Raman scattering by phonons is resonant. We find that the Raman spectra exhibit a crossover behavior for laser excitation close to the $\pi$-plasmon energy. This phenomenon is explained by the disappearance of certain paths for resonant Raman scattering and the appearance of other paths beyond this energy. Besides that, the electronic joint density of states is divergent at this energy, which is reflected on the behavior of the Raman bands of the two structures in a qualitatively different way. Additionally, a number of Raman bands, originating from divergent phonon density of states at the M point and at points, inside the Brillouin zone, is also predicted. The calculated spectra for graphene are in excellent agreement with available experimental data. The obtained Raman bands can be used for structural characterization of silicene and graphene samples by Raman spectroscopy.

\end{abstract}

\pacs{73.22.Pr, 63.22.Rc, 78.67.-n }

\vspace{2pc}
\noindent{\it Keywords}: graphene, silicene, Raman scattering

\maketitle

\section{Introduction}

Following the success of the synthesis of the atom-thin graphene, considerable efforts are currently directed towards the search of other two-dimensional structures \cite{butl13}. A planar honeycomb structure of silicon, termed silicene, has been proposed \cite{verr07}. In an ab-initio study \cite{caha09}, free-standing silicene has been relaxed to a slightly corrugated structure and its electronic and vibrational properties have been calculated. The intrinsic electrical transport of silicene has been examined from first-principles and the dominant role of the out-of-plane acoustic phonons in it has been identified \cite{li13}. Theoretical investigation of the electron-phonon coupling, limited to the in-plane optical phonons, has also been reported \cite{yan13}. 

Silicene sheets, deposited on silver substrate, have been characterized by scanning probe microscopy \cite{chen12,rest13} and angle-resolved photoemission spectroscopy \cite{vogt12,moll13}. It has been found that silicene forms a few commensurate phases with corrugated honeycomb structure.  The existence of linear electronic dispersion of silicene close to the Fermi energy has been established \cite{chen12,vogt12}. Strong hybridization between silicene and silver substrate has been evidenced by the collected characterization data \cite{maha14,cinq15} and theoretical studies \cite{maha14,cinq15,caha13}. While electron microscopy allows for the precise determination of the atomic structure of silicene, the identification of the various silicene phases can be done in a fast and nondestructive way by Raman spectroscopy. The measured Raman spectra normally show a single intense line, which is attributed to the optical bond-stretching mode $E_{2g}$ of silicene, as well as several less intense first- and second-order features \cite{scal14,cinq13}. The second-order bands have been found dependent on the laser excitation \cite{cinq13}. The effect of coverage, strain, charge doping, and defects on the Raman spectra of silicene phases, grown on different substrates, has been studied experimentally \cite{zhua15}. On the theoretical side, first-order Raman spectra of free-standing silicene and hydrogenated armchair silicene nanoribbons have been derived by convolution of the calculated vibrational spectrum \cite{scal12}. As far as we know, calculated second-order (defect-induced and two-phonon) bands of silicene have not been reported. While the calculation of these bands for silicene phases on a substrate constitutes a computational task of considerable complexity, the calculation of the two-phonon bands of free-standing silicene is feasible, as demonstrated in the case of the isostructural graphene \cite{popo12}. As it will be shown below, the two-phonon bands of free-standing silicene change their behavior on increasing the laser excitation across the $\pi$-plasmon energy. In this respect, it is instructive to study this behavior in comparison with that of graphene, for which experimental Raman data at laser excitations beyond the $\pi$-plasmon energy have only recently been reported \cite{tybxx}.

Here, we calculate the two-phonon Raman bands of free-standing silicene and graphene in a wide range of laser excitations within a density-functional non-orthogonal tight-binding (NTB) model. The paper is organized as follows. The theoretical background is described briefly in Sec. II. The results and discussion are presented in Sec. III. The paper ends up with conclusions (Sec. IV).

\section{Theoretical background}

The electronic band structure of silicene and graphene is obtained within the NTB model with Hamiltonian and overlap matrix elements taken over from previous \textit{ab-initio} studies on silicon dimers \cite{frau95} and carbon dimers \cite{pore95}. The tight-binding formalism is presented in sufficient detail elsewhere \cite{popo04} and will not be described here. The phonon dispersion is calculated using dynamical matrix, derived via a perturbative approach within the NTB model \cite{popo10}.  The summations over the Brillouin zone for the total energy and the dynamical matrix are performed over a $40\times40$ Monkhorst-Pack mesh, for which the lattice parameter and the phonon frequencies are converged to better than $0.01$ \AA~ and $1$ cm$^{-1}$, respectively.

The two-phonon Raman intensity is calculated within the NTB model \cite{popo12} using fourth-order quantum-mechanical perturbation theory, in which the electron-photon and electron-phonon Hamiltonians are considered as perturbations to the Hamiltonian of the non-interacting system of electrons, photons, and phonons. Each one of the fourth-order terms can be visualized as a sequence of virtual processes of absorption of an incident photon with creation of an electron-hole pair (state $a$), two consecutive electron/hole scattering processes by a phonon (states $b$ and $c$), and final electron-hole annihilation (state $f$). The two-phonon Raman intensity is given by the expression \cite{card75}
\begin{equation}
I\propto\sum_{f}\left|\sum_{c,b,a}\frac{M_{fc}M_{cb}M_{ba}M_{ai}}{\Delta E_{ic}\Delta E_{ib}\Delta E_{ia}}\right|^{2}\delta\left(E_{i}-E_{f}\right)\label{a2}
\end{equation}
where 
$\Delta E_{iu}=E_{i}-E_{u}-i\gamma$; the indices denote the initial ($i$), intermediate ($u=a,b,c$), and final ($f$) states of the system of photons, electrons, holes, and phonons. In the initial state, only an incident photon is present and $E_{i}=E_{L}$, where $E_{L}$ is the incident photon energy. We restrict ourselves to Stokes processes only. Then, in the final state, there is a scattered photon and two created phonons. $M_{uv}$ are the matrix elements for virtual processes between initial, intermediate, and final states. In particular, $M_{ai}$ and $M_{fc}$ are the matrix elements of momentum for processes of creation and recombination of an electron-hole pair, respectively.  $M_{ba}$ and $M_{cb}$ are the electron/hole-phonon matrix elements. The electron-photon and electron-phonon matrix elements are calculated explicitly \cite{popo05}. $\gamma$ is the sum of the halfwidths of pairs of electronic and hole states; it will be referred to as the \textit{electronic linewidth} and is also calculated here \cite{popo06a}. The Dirac delta function ensures energy conservation for the entire process. In the calculations, it is replaced by a Lorentzian with a half width at half maximum of $5$ cm$^{-1}$. The missing proportionality factor has a weak dependence on $E_L$, which is usually neglected \cite{card75}.

The summation over the intermediate states runs over all valence and conduction bands, and over all electron wavevectors $\mathbf{k}$. The summation over the final states runs over all phonon branches and phonon wavevectors $\mathbf{q}$. For both summations, convergence is reached with a $400\times400$ mesh of $\mathbf{k}$ and $\mathbf{q}$ points in the first Brillouin zone. All eight possible resonant scattering processes, composed of the mentioned virtual processes, are taken into account \cite{popo12}. Backscattering geometry is implied.

Equation 1 applies only for Raman scattering under resonant conditions, i.e., when the laser excitation energy matches transitions of the system. The Raman intensity is then enhanced because the real part of one, two, or three of the factors $\Delta E_{iu}$ vanishes. This correlates the laser excitation energy, the initial and final energy of the scattered electron/hole, and the energy of the scattering phonons. Generally, this leads to \textit{dispersive} two-phonon Raman bands, i.e., to dependence of their position on the excitation energy \cite{thom00}. Following the tradition, we call such resonant processes and the originating bands \textit{double-resonant} (DR). 
As we shall see, enhancement of the Raman intensity can also take place for scattering phonons close to points with divergent two-phonon density of states (DOS). The position of such two-phonon bands is determined by the phonon energy at such points \cite{vanh53} and, therefore, the bands are \textit{non-dispersive}. In both cases, in order for a Raman band to be observed, it is also necessary that the electron-phonon coupling is relatively large for one or both phonons, taking part in the two-phonon scattering process. 
 
\begin{figure}[tbph]
\includegraphics[width=85mm]{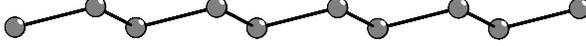} 
\caption{Sideview of the relaxed atomic structure of silicene. The structure is corrugated with out-of-plane corrugation of $0.57$~\AA.}
\end{figure}

\section{Results and Discussion}

The hexagonal honeycomb structure of silicene and graphene is relaxed within the NTB model \cite{popo04}. This model predicts planar structure of graphene \cite{popo12} and corrugated one of silicene (Fig. 1). The out-of-plane corrugation of 0.57 \AA~is larger than the \textit{ab-initio} value of 0.44 \AA, while the relaxed lattice parameter of 3.83 \AA~ is equal to the \textit{ab-initio} result \cite{caha09,li13,yan13,scal12}. The corrugation of silicene is essential for stabilizing the lattice and avoiding imaginary phonon frequencies of the out-of-plane acoustic phonons. The experimentally observed silicene on Ag(111) has out-of-plane corrugation in the range \cite{scal14} $0.7-1.0$ \AA, which is more than twice larger than that estimated one for free-standing silicene.

\begin{figure}[tbph]
\includegraphics[width=85mm]{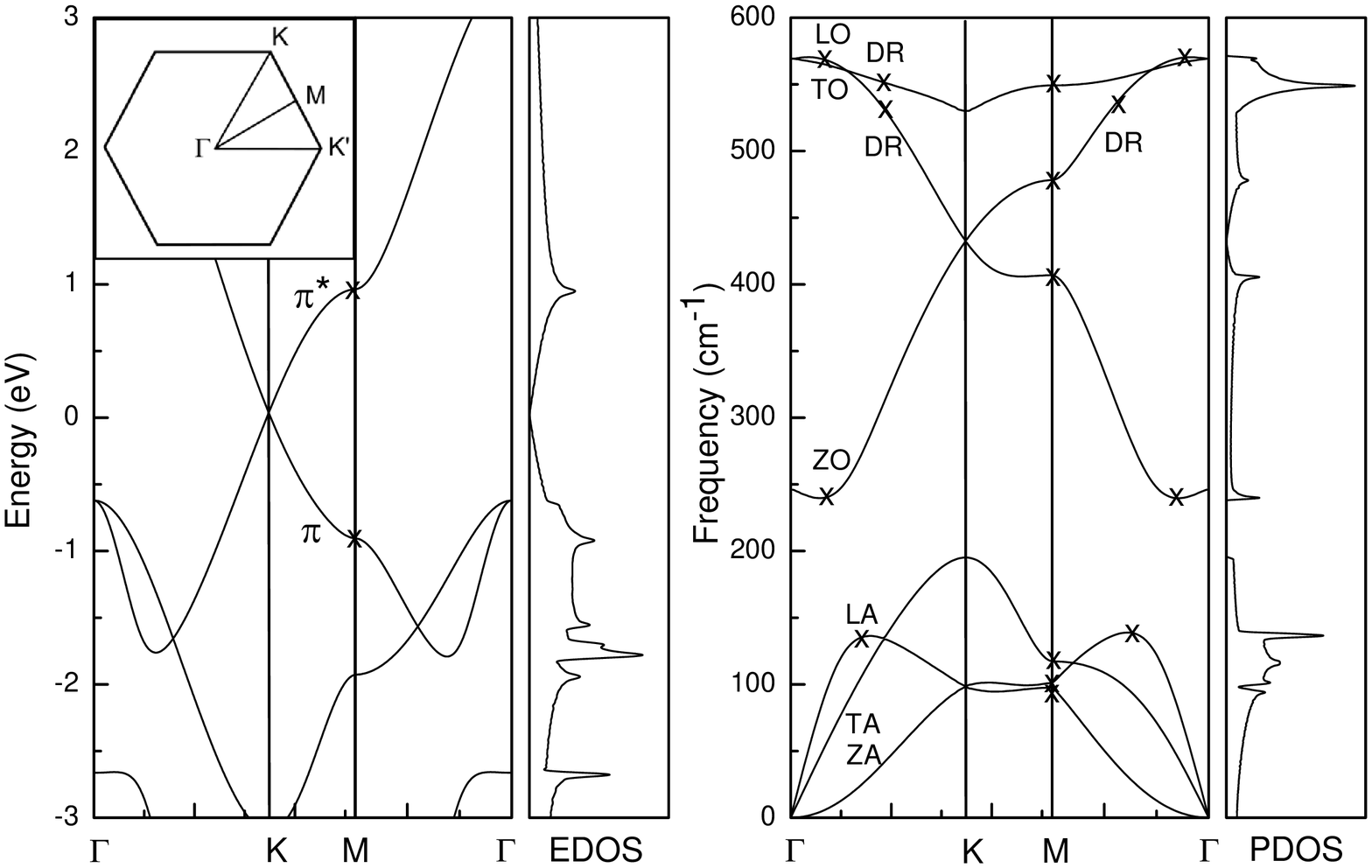} 
\caption{Left: electronic band structure of silicene in the vicinity of the Fermi energy, taken for zero, and corresponding electronic DOS. The saddle points of the valence $\pi$ and conduction $\pi^{*}$ bands are marked by crosses. Inset: Brillouin zone with special points. Right: phonon dispersion of silicene and corresponding phonon DOS. The phonon branches are denoted by acronyms, connected to the character of the phonon eigenvectors (see text). The phonons at phonon DOS singularities with significant contribution to the two-phonon Raman spectrum are marked by crosses and those, contributing to DR processes, are marked by crosses and the symbol $^{''}$DR$^{''}$.}
\end{figure}

The calculated electronic band structure of silicene, similarly to graphene \cite{popo12}, has linear valence $\pi$ and conduction $\pi^{*}$ bands close to the Fermi energy (Fig. 2, left). These bands have saddle points at the M point with logarithmic divergent electronic DOS \cite{vanh53}, separated at the $\pi$-plasmon energy $E_p=1.83$ eV. This value is lower than the NTB one for graphene $E_p=5.04$ eV. The $\pi^{*}$ and valence bands agree fairly well with the \textit{ab-initio} ones, while the higher-energy conduction bands are less well reproduced \cite{caha09,li13,yan13,scal12}. The electronic DOS has also other saddle-point singularities at the M point and at points, inside the Brillouin zone (Fig. 2, left). Among the various singularities of the electronic DOS, those at the saddle points of the $\pi$ and  $\pi^{*}$ bands play important role in the enhancement of the two-phonon Raman intensity.

The derived phonon frequencies of silicene overestimate the \textit{ab-initio} values \cite{caha09}. However, fair agreement can be achieved by downscaling the former by a factor of $0.83$ (Fig. 2, right). Such an overestimation is also observed in the tight-binding phonon dispersion of graphene, where the required scaling factor is $0.9$ \cite{popo06}. The phonon branches are denoted by the symbols of the $\Gamma$-point phonons, describing their atomic displacement pattern. Each symbol consists of two letters with the meaning: O (optical), A (acoustic), L (in-plane longitudinal), T (in-plane transverse), and Z (out-of-plane transverse). The phonon DOS has similar divergent singularities as the electronic DOS (Fig. 2, right). As it will be shown below, the phonon DOS singularities yield specific enhancement of the two-phonon Raman bands.

\begin{figure}[tbph]
\includegraphics[width=80mm]{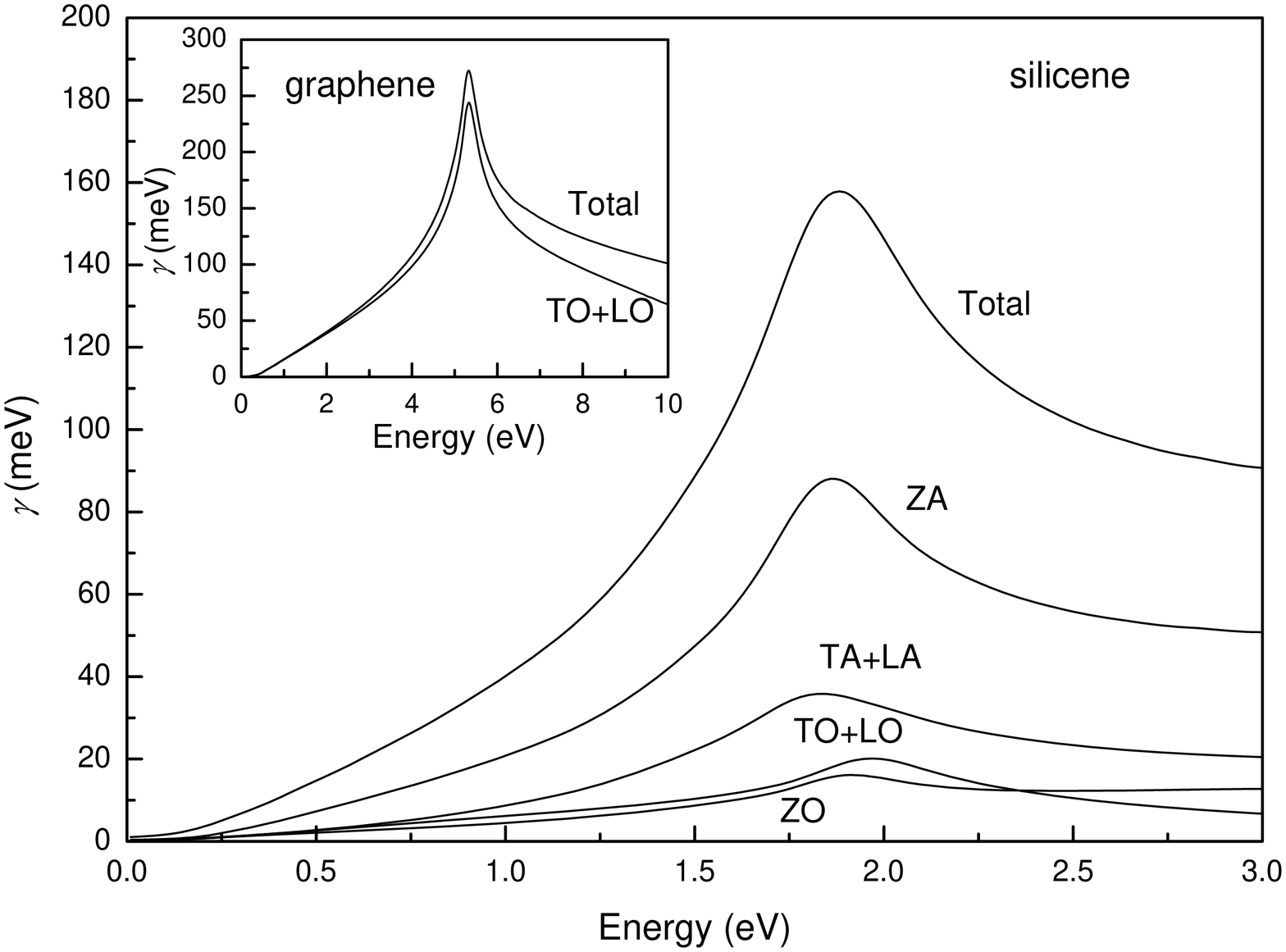} 
\caption{Energy dependence of the electronic linewidth of silicene. The ZA phonons have largest contribution to the linewidth. Inset: electronic linewidth of graphene. The TO+LO phonons play dominant role in the electron scattering in graphene.  For both silicene and graphene, the maximum of the linewidth  corresponds to the $\pi$-plasmon energy $E_p$.}
\end{figure}

The  two-phonon Raman intensity, Eq. 1, includes the electronic linewidth $\gamma$. Normally, the electron-phonon scattering provides the dominant contribution to it in perfect single-layer periodic structures (see, e.g., Ref.~\cite{popo12} and references therein). In silicene, as it is seen in Fig. 3, the electron-phonon scattering contribution to the electronic linewidth increases for phonons in the following order: ZO and TO+LO, TA+LA, and ZA, in agreement with previous \textit{ab-initio} results \cite{li13}. Such a behavior is quite distinct from that in graphene, where the TO+LO phonons have major contribution to the linewidth (Fig. 3, inset). The enhanced electron scattering by ZA phonons in silicene can be attributed to the corrugated atomic structure. It is expected to result in a considerable reduction of the electron scattering time and to worsening the characteristics of the electronic transport in silicene compared to those of graphene. 

\begin{figure}[tbph]
\includegraphics[width=80mm]{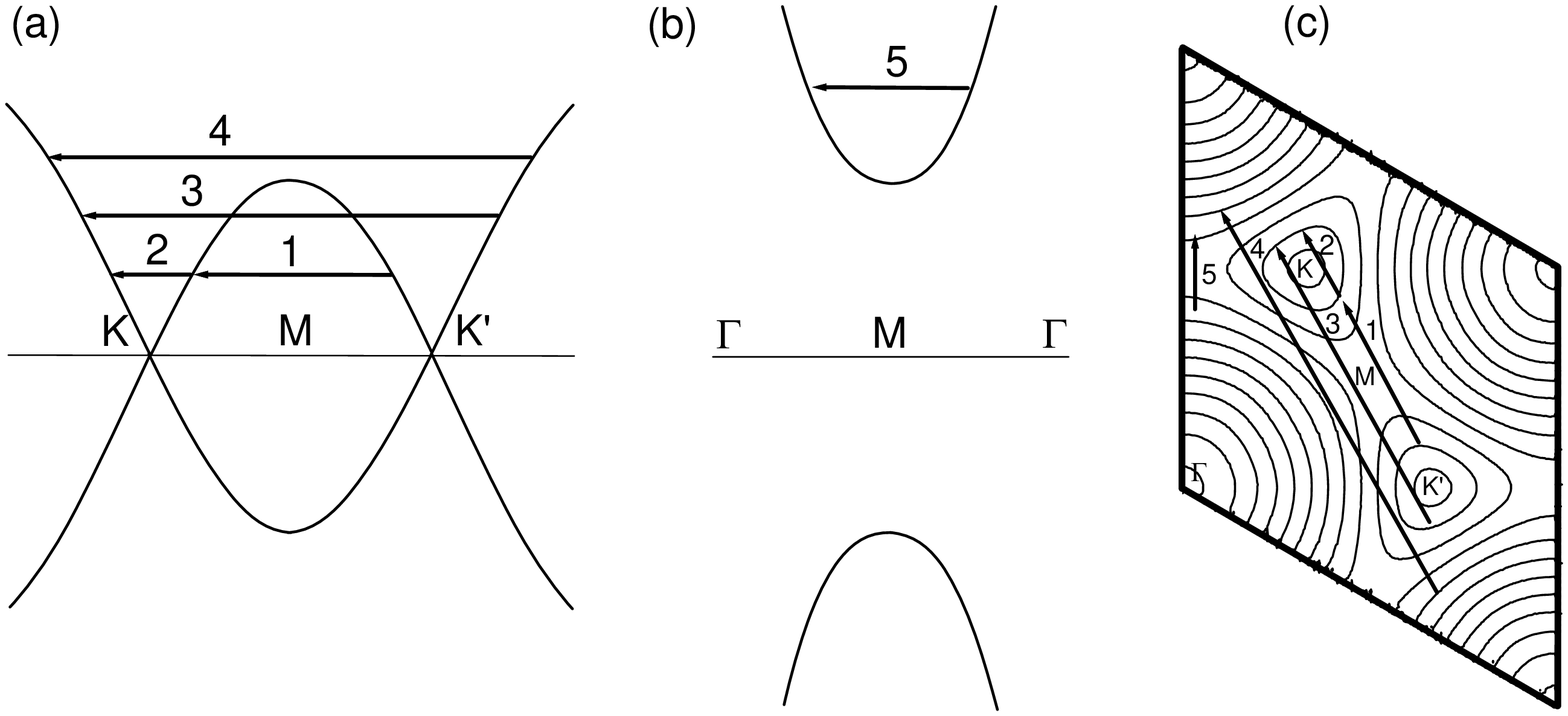} 
\caption{(a) Schematic electronic bands close to the Fermi energy along the KMK$^{'}$ direction. (b) Same along the  $\Gamma$M$\Gamma$ direction. (c) Equi-energy contour map of the $\pi^{*}$ band in the entire Brillouin zone. In all three plots, the arrows are examples of phonon wavevectors for scattering at energies below $E_p$: $^{''}$inner$^{''}$ and $^{''}$outer$^{''}$ intervalley processes (1 and 3, respectively) and intravalley processes (2), and at energies above $E_p$: $^{''}$outer$^{''}$ intervalley processes (4) and intravalley processes (5). The contribution of $^{''}$outer$^{''}$ processes 3 and 4 to the Raman intensity is usually small.}
\end{figure}

Before proceeding with the calculated Raman spectra of silicene, we first discuss the two-phonon spectra of graphene at laser excitation in the visible range \cite{popo12}. The electronic band structure of graphene, featuring Dirac cones, favors DR scattering processes. Indeed, there are always pairs of electronic states at conduction and valence Dirac cones, matching the laser excitation for direct electronic transitions with creation of electron-hole pairs and yielding vanishing difference $E_i - E_a$. Additionally, there are always phonons for  intravalley or intervalley scattering of electrons or holes (Figs. 4(a) and 4(c)), such that $E_i - E_b$ is vanishingly small. Finally, the scattering of holes or electrons by phonons with  momenta, opposite to the previous ones, yields small $E_i - E_c$. Therefore, single, double, or triple resonances can take place, the latter giving dominant contribution to the Raman intensity. Moreover, the position of the originating two-phonon Raman bands will depend on the laser excitation, i.e., they are dispersive.

All derived Raman bands of graphene for visible laser excitation are enhanced exclusively through the DR mechanism and no measurable Raman signal is associated with singularities of the two-phonon DOS. The most intense 2D band comes from intervalley scattering by TO phonons close to the K point. The less intense 2D$^{'}$ band arises from intravalley scattering by LO phonons close to the $\Gamma$ point. The two bands are well-separated because of the steep electronic dispersion at the K point and the relatively large dispersion of the LO and TO phonons. The major contribution to the 2D band comes from scattering processes between electronic states within the segment KMK$^{'}$, so-called $^{''}$inner$^{''}$ processes, while scattering processes outside KMK$^{'}$, so-called $^{''}$outer$^{''}$ processes, have much smaller contribution. Other, less intense Raman bands, due to intravalley and intervalley scattering by other phonons are also predicted and observed. 

\begin{figure}[tbph]
\includegraphics[width=85mm]{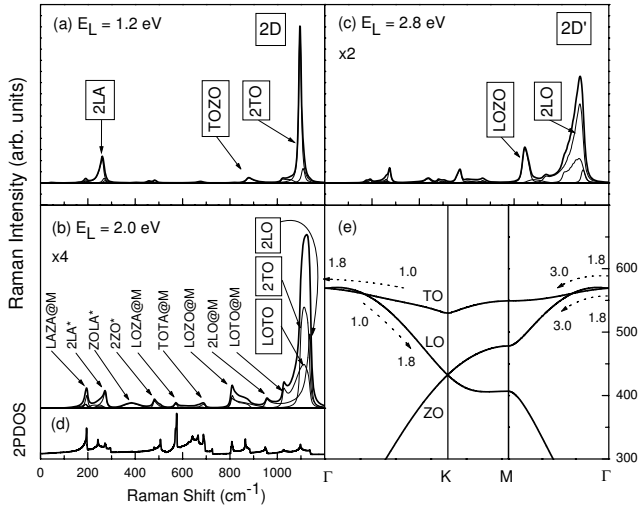} 
\caption{Two-phonon Raman spectrum of silicene (thick line) and contributions to the spectrum by various pairs of phonon branches (thin lines) at laser excitation (a) $E_L=1.2$ eV,  (b) $E_L=2.0$ eV, and (c) $E_L=2.8$ eV. The most prominent two-phonon DR modes are marked by the acronyms of the two phonons. The additional sybmbols $^{''}$M$^{''}$ and $^{''}$*$^{''}$ denote two-phonon modes, arising from M point phonons and phonons inside the Brillouin zone, respectively (see Fig. 2). The most intense Raman bands originate from DR overtone and combination modes of LO and TO phonons. (d) Two-phonon DOS of silicene. (e) Part of the phonon dispersion of silicene. The wavevectors for DR processes at $E_L$ between $1$ and $3$ eV are shown by dotted arrows.}
\end{figure}

Next, we turn to the calculation of the two-phonon Raman spectra of silicene.  The spectrum at $E_L=1.2$ eV (Fig. 5(a)) has three more prominent bands, which are due to DR processes. The intense one  at about $1100$ cm$^{-1}$ is the 2D band, which consists mainly of the 2TO mode. This mode arises from $^{''}$inner$^{''}$ scattering by TO phonons with wavevectors of length of about two-thirds of the $\Gamma$K distance (Fig. 5(e)), as evidenced from the careful analysis of the wavevectors of the contributing phonons. The 2LO mode is weak and overlaps with the 2D band and, therefore, no distinct 2D$^{'}$ band should be observed. The 2LA and TOZO bands are comparatively weak. With increasing $E_L$, the wavevector of the TO phonons, contributing to the 2D band, decreases until reaching length zero at laser excitations, close to $E_p$. This is accompanied by a significant decrease of the 2D band intensity. The wavevectors of the LO phonons, contributing to the LO mode, increase in length from about one fourth to about two-thirds of  the $\Gamma$K distance, while the intensity of this mode increases.

At $E_L=2.0$ eV, slightly above $E_p$, the Raman spectrum has an intense DR band at about $1100$ cm$^{-1}$ and a number of weak features (Fig. 5(b)). The intense band comprises the overtone modes 2TO and 2LO, and the combination mode LOTO, which are of comparable intensity and are enhanced by scattering processes of type $^{''}$$5$$^{''}$ (Fig. 4(b) and 4(c)). 

The weak bands in the spectrum at $E_L=2.0$ eV can be attributed to singularities of the two-phonon DOS and to non-negligible electron-phonon coupling for all phonon branches. The role of the two-phonon DOS in the enhancement of the Raman signal can be elucidated by analyzing Eq. 1. In the crude approximation of wavevector-independent addends at a given $E_{L}$, the sum over the phonons in the entire Brillouin zone is essentially reduced to the two-phonon DOS \cite{born54}. In two-dimesional structures, the latter is known to diverge logarithmically at saddle-type critical points or to have quasi-onedimensional singularities at points, inside the BZ \cite{vanh53}. Therefore, the two-phonon Raman intensity is expected to show similar behavior in the vicinity of such points. This correlation between the divergent two-phonon DOS features and the two-phonon Raman bands is clearly seen in Fig. 5(b) and 5(d). Most of these bands can be connected to singularities of the two-phonon DOS at the M point of the Brillouin zone. The weak bands 2LA$^{*}$, ZOLA$^{*}$, and 2ZO$^{*}$ originate from phonons close to the maximum and minimum of the LA and ZO branches, where the two-phonon DOS has quasi-onedimensional singularity. There is a contribution to the Raman intensity from phonons at the maximum ($^{''}$overbending$^{''}$) of the LO branch but it overlaps with the 2LO mode. The intensity of the bands depends on the characteristics of the singularities and on the electron-phonon coupling. Two-phonon Raman bands, enhanced by divergent two-phonon singularities, have been predicted and discussed in bilayer graphene \cite{popo15}. It can also be noticed that non-divergent features of the two-phonon DOS do not bring about noticeable Raman bands. 

At $E_L=2.8$ eV, much above $E_p$, several DR bands are predicted. The intense band at $1100$ cm$^{-1}$ is the 2D$^{'}$ band (Fig. 5(c)). It is formed mainly from the 2LO mode. Thus, the 2D band transforms gradually into the 2D$^{'}$ band with increasing $E_L$ past the $\pi$-plasmon energy.

\begin{figure}[tbph]
\includegraphics[width=80mm]{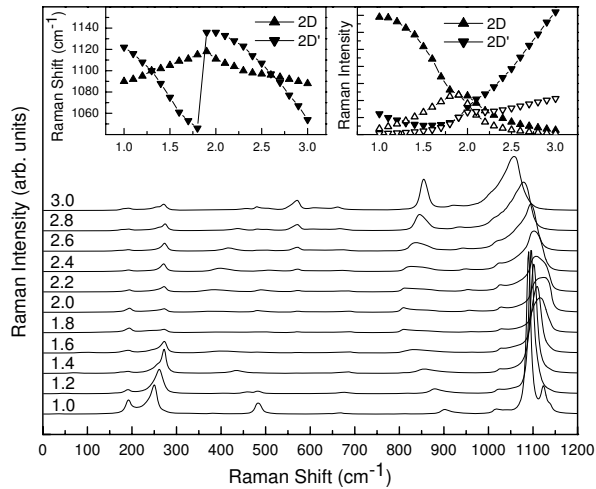} 
\caption{Raman spectra of silicene at laser excitation $E_L$ from $1.0$ to $3.0$ eV. The spectra are displaced vertically for better visibility. Left inset: the dependence of the Raman shift of the 2D and 2D$^{'}$ bands on $E_L$. Right inset: the dependence of the integrated Raman intensity of the 2D and 2D$^{'}$ bands on $E_L$. The empty symbols are for fixed linewidth $\gamma=160$ meV. The lines are guides to the eye. }
\end{figure}

The evolution of the Raman bands of silicene with increasing $E_L$ is shown in Fig. 6. The DR bands have a clear dependence on $E_L$, i.e., they are dispersive, while the bands, due to singularities of the two-phonon DOS, are non-dispersive. Although the dependence of the DR bands position on the laser excitation is non-linear (Fig. 6, left inset), it can be described approximately by the slope of the curves, so-called \textit{dispersion rate}. The derived rate is  $30$ cm$^{-1}$/eV for the 2D band below $1.8$ eV and $-100$ cm$^{-1}$/eV for the 2D$^{'}$ band above $2.2$ eV. The change of the Raman shift and intensity with $E_L$ (Fig. 6, left and right insets) is consistent with the change of the electron scattering paths across the $\pi$-plasmon energy.

\begin{figure}[tbph]
\includegraphics[width=85mm]{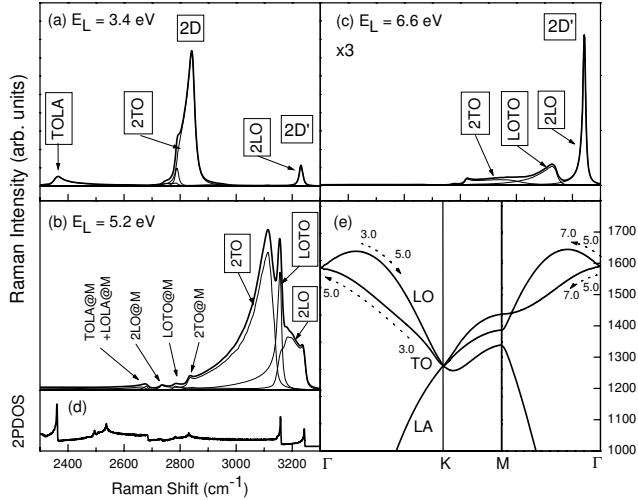} 
\caption{Two-phonon Raman spectrum of graphene (thick line) and contributions to the spectrum by various pairs of phonon branches (thin lines) at laser excitation (a) $E_L=3.4$ eV,  (b) $E_L=5.2$ eV, and (c) $E_L=6.6$ eV. The most intense Raman bands originate from DR overtone and combination modes of LO and TO phonons. (d) Two-phonon DOS of graphene. (e) Part of the phonon dispersion of graphene. The wavevectors for DR processes at energies between $3$ and $7$ eV are shown by dotted arrows.}
\end{figure}

We perform similar calculations of the two-phonon Raman spectrum of graphene. For comparison with the spectrum of silicene, we extend the previous calculations on graphene \cite{popo12} in the ultraviolet range up to $7.0$ eV, above $E_p$. At $E_L=3.4$ eV (Fig. 7(a)), the Raman spectrum exhibits distinct DR features in the range $2300-3300$ cm$^{-1}$: the intense 2D band and the weak bands 2D$^{'}$ and D+D$^{''}$ (or TOLA band). The 2D band originates from scattering by TO phonons with wavevectors, close to one half of the $\Gamma$K distance (Fig. 7(e)). With increasing $E_L$ up to about $5.0$ eV, the length of the wavevector of the scattering TO phonons decreases to zero, accompanied by a significant widening of the 2D band, while its peak intensity shows only a small increase. Similar behavior has the TOLA band but with participation of TO and LA phonons. The 2D$^{'}$ band is due to scattering by LO phonons with wavevectors close to one half of the $\Gamma$K distance. With increasing $E_L$ up to about $5.0$ eV, the length of the wavevectors of the scattering LO phonons increase to about two-thirds of $\Gamma$K and the 2D$^{'}$ band intensity increases. 

At $E_L=5.2$ eV, slightly above $E_p$, the calculated Raman spectrum shows a broad asymmetric band, mostly of DR origin (Fig. 7(b)). As for silicene, for energies above $E_p$,  $^{''}$inner$^{''}$ scattering by TO phonons is no longer possible. The principal scattering path for LO and TO phonons is along the $\Gamma$M$\Gamma$ direction (Fig. 7(e)). With increasing $E_L$, the LO and TO phonon wavevectors for such processes increase from zero, accompanied by decrease of the peak intensity of the 2D and 2D$^{'}$ bands. The asymmetric band has a characteristic sharp peak of the LOTO mode, which arises from scattering by pairs of LO and TO phonons close to the $\Gamma$ point.  

Besides the intense DR band, the Raman spectrum  at $5.2$ eV has several weak bands due to divergent two-phonon DOS at the M point (bands LOLA@M, TOLA@M, 2LO@M, LOTO@M, and 2TO@M) and at the overbending of the LO branch (kink at about $3240$ cm$^{-1}$). The predicted spectrum is in excellent agreement with recent Raman data \cite{tybxx}. Finally, the spectrum at $E_L=6.6$ eV has only DR bands, namely, an intense 2LO band (or 2D$^{'}$ band) and weak bands 2TO and LOTO (Fig. 7(c)).

\begin{figure}[tbph]
\includegraphics[width=80mm]{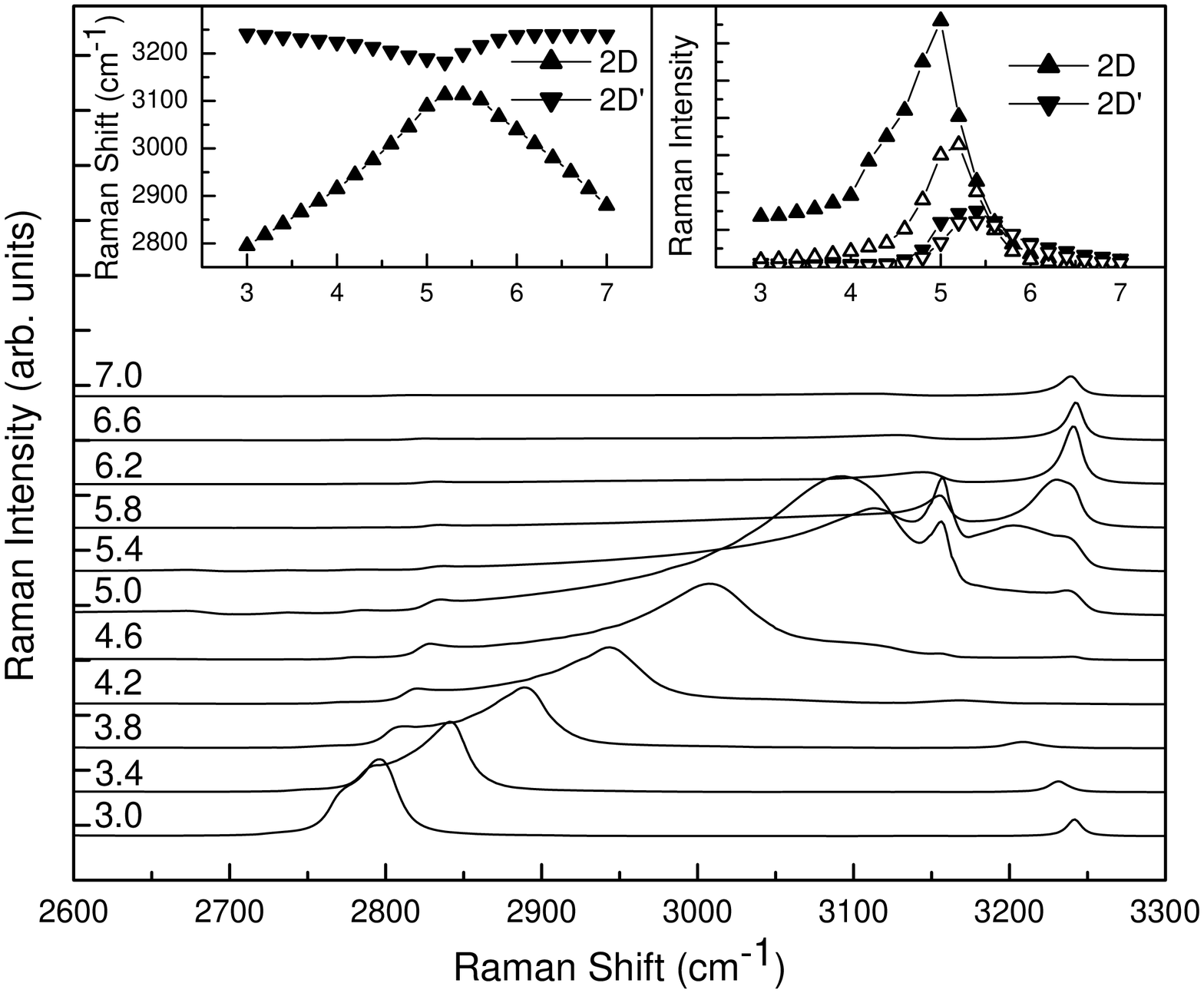} 
\caption{Raman spectra of graphene at laser excitation $E_L$ from $3.0$ to $7.0$ eV. The spectra are displaced vertically for better visibility. Left inset: the dependence of the Raman shift of the 2D and 2D$^{'}$ bands on $E_L$. Right inset: the dependence of the integrated Raman intensity of the 2D and 2D$^{'}$ bands on $E_L$. The empty symbols are for fixed linewidth $\gamma=260$ meV. The lines are guides to the eye. }
\end{figure}

The calculated two-phonon Raman spectra at laser excitation in the range $3.0-7.0$ eV (Fig. 8) show gradual increase of the 2D and 2D$^{'}$ band intensity in the vicinity of $E_p$. Both bands are dispersive with dispersion rates of $150$ cm$^{-1}$/eV for the 2D band below $5.0$ eV and $40$ cm$^{-1}$/eV for the 2D$^{'}$ band above $5.2$ eV (Fig. 8, left inset). The intensity of both bands increases steeply in the vicinity of $E_p$ (Fig. 8, right inset), which corresponds to the saddle-point singularity of the joint DOS.

The evolution of the 2D and 2D$^{'}$ bands of silicene and graphene with the laser excitation essentially depends on the electronic linewidth. In the case of constant electronic linewidth, the Raman bands in both structures mimic the joint DOS. The latter brings about a sharp increase of the integrated intensity of the 2D and 2D$^{'}$ bands of graphene and silicene on approaching $E_p$. The use of energy-dependent linewidth for graphene modifies only quantitatively the singularity of the intensity of the two bands at $E_p$ (Fig. 8, right inset). However, it has a significant effect on both bands in silicene, where the singularity of the intensity at $E_p$ is modified to a sharp increase for the 2D$^{'}$ band and to a sharp decrease for the 2D band with increasing $E_L$ across $E_p$ (Fig. 6, right inset). Such a behavior has been observed experimentally in silicene samples \cite{cinq13}. The relative increase of the intensity away from $E_p$ is a consequence of the decrease of the linewidth for such energies, as seen from Eq. 1 The results for both materials show that the behavior of the Raman intensity depends on the interplay of the joint DOS and the electronic linewidth. While in graphene the effect of the former prevails, in silicene the evolution of the bands with the laser excitation is essentially modified, when energy-dependent linewidth is used. This result underlines the importance of taking into account the energy dependence of the electronic linewidth in the calculation of the resonant Raman spectra.

\section{Conclusions}

We have presented a computational study of the two-phonon Raman spectra of silicene and graphene within a density-functional non-orthogonal tight-binding model. We have found that the evolution of the Raman bands with the laser excitation  in the vicinity of the $\pi$-plasmon energy is determined by specific scattering paths. In the same energy region, the behavior of the spectra  is found to mimic the joint DOS. This behavior is affected to a different extent by the use of energy-dependent electronic linewidth instead of a constant one. Namely, the spectra of graphene are only quantitatively modified by the latter substitution, while the spectra of silicene are drastically changed. A number of Raman features, enhanced by divergent two-phonon DOS have also been identified in both structures. The obtained spectra for graphene are in excellent agreement with recent Raman data. Our study emphasizes the necessity of explicitly accounting of the electronic band structure, phonon dispersion, and electronic linewidth for the prediction of the resonant Raman intensity.  

\ack

V.N.P. acknowledges financial support from EU Seventh Framework Programme REGPOT Project 316309 INERA.

\section*{References}


\providecommand{\newblock}{}

\end{document}